# Computational modelling in single-cell cancer genomics: methods and future directions

Allen W Zhang[1] and Kieran R Campbell[2,3,4]

## Abstract

Single-cell technologies have revolutionized biomedical research by enabling scalable measurement of the genome, transcriptome, and proteome of multiple systems at single-cell resolution. Now widely applied to cancer models, these assays offer new insights into tumour heterogeneity, which underlies cancer initiation, progression, and relapse. However, the large quantities of high-dimensional, noisy data produced by single-cell assays can complicate data analysis, obscuring biological signals with technical artefacts. In this review article, we outline the major challenges in analyzing single-cell cancer genomics data and survey the current computational tools available to tackle these. We further outline unsolved problems that we consider major opportunities for future methods development to help interpret the vast quantities of data being generated.

## Introduction

Cancers are invasive neoplasms derived from single cells that undergo successive acquisition of cell-specific properties through somatic mutation. The clonal diversity that results can lead to metastasis, histologic transformation, and treatment resistance (Greaves & Maley 2012). The trellis of phenotypes generated from this process of branching evolution is subject to the selective pressures of the *tumour microenvironment* (TME, see **Table 1**, McGranahan et al. 2017; Zhang et al. 2018). High-throughput approaches such as bulk whole-genome sequencing, RNA-seq, flow cytometry, mass cytometry, and immunohistochemistry have been extensively applied to establish the foundations of our current understanding of carcinogenesis and the cancer cell-microenvironment interface (Weinstein et al. 2013; Shah et al. 2012; Curtis et al. 2012). However, these approaches are generally limited by tradeoffs between the breadth and granularity of genotypic and phenotypic information obtained.

Recent technological advances have enabled multimodal profiling of the genomes, transcriptomes, and proteomes from thousands to millions of single cells. The advent of single-cell RNA-sequencing (scRNA-seq) at scale has resulted in an explosion of recent studies capturing not only malignant phenotypes but also cellular states of immune and stromal cells in the tumour microenvironment (A. W. Zhang et al. 2019). Additionally, advances in DNA sequencing technologies that allow for single-cell measurements of point mutations or large scale copy number aberrations have enabled the clonal decomposition of primary tumours, cell lines, and patient-derived xenografts in unprecedented detail, leading to new insights into aneuploidy (Laks et al. 2019), oncogenic processes (Yu et al. 2014), and treatment resistance (Chen et al. 2018; Kim et al. 2018). These same methods can be used to detect and functionally characterize circulating tumour cells (CTCs), which have the potential to be used as early diagnostic markers or markers of metastasis. Similarly, single-cell proteomic technologies such as mass cytometry allow scalable measurement of multiple protein markers while retaining information about the

1. MD/PhD Program, Faculty of Medicine, University of British Columbia, 317-2194 Health Sciences Mall, Vancouver, BC, Canada
2. Lunenfeld-Tanenbaum Research Institute of Sinai Health System, 600 University Ave, Toronto, ON, Canada
3. Department of Molecular Genetics, University of Toronto, Medical Science Building, 1 King's College Cir, Toronto, ON, Canada
4. kierancampbell@lunenfeld.ca

spatial origins of cells, revealing the high-dimensional architecture of cancers (Jackson et al. 2020).

However, the emergence of these technologies has created a deluge of noisy, high-dimensional data, the analysis and interpretation of which is key to understanding cancer pathogenesis and aetiology. To address this, a range of computational methods have been developed specifically for the analysis of cancer genomics data. In this review, we outline the challenges and methods associated with the computational analysis of single-cell genomic, transcriptomic and proteomic data in the context of cancer research (**Figure 1a-b**). We further point to unsolved problems and future directions in the development of computational tools for single-cell cancer genomics, concerning (i) inferring clonal dynamics over time and space, (ii) understanding the impact of TME heterogeneity on cancer evolution, (iii) machine learning approaches to make single-cell assays predictive of therapeutic response, and (iv) methods to uncover interactions and signatures from spatially resolved data. We intend this review to serve as a point of reference of the current state of the field as well as an opportunity to encourage discussion on future computational methods necessary to realize the potential of single-cell cancer genomics.

## Computational methods for single-cell cancer genomics

### Mutation profiling and phylogenetic inference

The noise inherent in detecting mutation events from picograms of DNA at the single-cell level — both point mutations and copy number aberrations — has led to a number of methods for mutation identification and the subsequent clustering of cells into *clones*. SNV detection from targeted single-cell DNA sequencing is complicated by high false negative rates due to amplification failures at heterozygous variants. To address this, probabilistic models such as Monovar (Zafar et al. 2016) and the Single Cell Genotyper (Roth et al. 2016) have been developed to assign variants by pooling strength across multiple cells. Monovar assumes each locus is independent and calls global variants present in any cell before re-computing cell-specific variant probabilities. In contrast, the Single Cell Genotyper models population structure by assuming cells belong to discrete clusters or clones, and uses a variational Bayes approach to compute the posterior probabilities of single-cell genotypes and cluster assignments, allowing for potential doublets.

Another set of methods performs CNV detection from near-uniform coverage single-cell whole genome sequencing (WGS) data. Preamplification-free approaches, such as direct library preparation (DLP) (Zahn et al. 2017) and DLP+ (Laks et al. 2019) produce low depth WGS data (< ~0.05X) with low amplification bias compared to alternative approaches (Hellani et al. 2004), making them amenable to copy number inference but not single-cell level variant calling. To analyze the outputs of these assays, existing methods that have been applied to bulk WGS data such as HMMCopy (Ha et al. 2012) as well as newer single-cell specific methods that leverage information from multiple cells simultaneously CHISEL (Zaccaria & Raphael 2019) have been used. HMMcopy computes copy number profiles based on a range of *ploidy* assumptions while correcting for GC content and mappability effects, and then selects the solution that minimizes non-integer copy number predictions. CHISEL uses *B-allele frequency* information to compute allele- and haplotype-specific copy number, which can be applied to identify cancer-associated loss-of-heterozygosity events (McGranahan et al. 2017).

Once single-cell genotypes have been defined, a common secondary analysis is to find a mutation tree that explains the observed data well — a process known as *phylogenetic inference* — for which a multitude of methods have been proposed. OncoNEM (Ross & Markowetz 2016) uses a probabilistic scoring function to measure how well a given tree fits observed mutation data accounting for false positive and negative rates under the *infinite sites assumption*, and searches for an optimal tree configuration using a heuristic local search procedure. SCITE (Jahn et al. 2016) similarly employs a probabilistic model under the infinite sites assumption, but uses *Markov Chain Monte Carlo* (MCMC) to sample trees from the posterior probability distribution using prune and reattach moves. Other phylogenetic models have been created to relax some modelling assumptions in previous work, such as SiFit (Zafar et al. 2017), which extends the OncoNEM and SCITE to allow for violations of the infinite sites assumption and consequently models recurrent evolution at each site. Refining this idea, Scarlett (Satas et al. 2019) only allows for mutation losses in regions for which there is evidence of copy number losses to account for the fact that mutation loss in other regions (with neutral copy number or copy number gains) is highly unlikely. Further methods in this space allow for more complex experimental designs, such as CALDER (Myers et al. 2019), which reconstructs phylogenetic trees from longitudinally acquired samples and builds in temporal constraints to improve the accuracy of the inferred trees. Finally, this inherent tree structure of single-cell genomic data may be exploited to refine single-cell variant calling using methods such as SCIΦ (Singer et al. 2018).

| | |
|---|---|
| *Tumour microenvironment* | The collection of immune, stromal, and vascular cells that may surround or infiltrate a tumour |
| *Ploidy* | The number of sets of chromosomes in a cell |
| *B-allele frequency* | Ratio of intensities between two alleles at a specified locus (heterozygous diploid = 0.5) |
| *Phylogenetic inference* | Reconstruction of the ancestral mutation tree that gave rise to the observed mutational profile in a tumour |
| *Clone* | Group of tumour cells that share a similar mutation profile |
| *Infinite sites assumption* | A given site or loci will be mutated at most once in the life history of a tumour |
| *Flow cytometry* | A method to measure physical properties of cells including protein expression via the scattering of light |
| *Overfitting* | Phenomena in a statistical or machine learning model where parameter estimates or predictions are overly influenced by the training dataset used and do not generalize well to new datasets |
| *Copy number aberrations* | Regions of the genome that have been amplified or depleted due to double strand breakage |
| *Markov Chain Monte Carlo* | A set of algorithms to sample from a probability distribution that is impossible to calculate exactly |
| *Approximate Bayesian Computation* | A method of parameter inference when the probability of the data given parameters is difficult to compute, but data can be simulated given the parameters |

**Table 1:** definitions of major concepts relevant to computational modelling of single-cell cancer genomics data.

### Gene expression

The advent of high-throughput single-cell RNA sequencing methods have ushered an era of intense computational methods development over the last 5 years. Essential analytical elements — quality control (McCarthy et al. 2017), dimensionality reduction (Pierson & Yau 2015; Risso et al. 2018), and clustering (Kiselev et al. 2017; Duò et al. 2018) — were the focus of initial studies, and are covered in other reviews (Rostom et al. 2017). Early single cell RNA-seq cancer studies leveraged these techniques to determine cell type composition (Li et al. 2017), intratumoural heterogeneity (Chen et al. 2018), and signatures of therapeutic resistance (Kim et al. 2018).

However, some of these elements necessitate revisiting in the context of cancer samples. Many pioneering single-cell RNA-sequencing studies used largely homogeneous cell lines as the substrate, and thresholds on mitochondrial gene percentage and transcribed gene count used to filter out low quality cells were designed for those scenarios (Ilicic et al. 2016; O'Flanagan et al. 2019). In contrast, cancer cells are associated with higher mitochondrial content (Osorio & Cai 2020), and the process of mechanical or enzymatic disaggregation of solid tumour samples prior to library preparation may result in increased expression of cellular stress markers, including mitochondrial genes. Additionally, plasma cells in the tumour microenvironment express a restricted profile of genes dominated by immunoglobulins, and are subject to being filtered out by transcribed gene thresholds.

Beyond preprocessing, cell type classification enables a broad survey of the tumour microenvironment. Many analyses (Shih et al. 2018; Kim et al. 2018) employ *ad hoc* approaches following unsupervised clustering to assign clusters to cell types. Bespoke approaches for cell type assignment that perform comparisons to bulk or single-cell RNA-seq data of purified populations (either experimentally or *in silico*) have also been developed (Zheng et al. 2017), including correlation-based methods such as scmap (Kiselev et al. 2018) and SingleR (Aran et al. 2019). However, in the context of perturbed cell states in the tumour microenvironment, data for comparable purified populations is often unavailable or not reflective of altered expression patterns (Shiga et al. 2015; Wherry 2011). To solve this problem, several approaches such as CellAssign (A. W. Zhang et al. 2019), Garnett (Pliner et al. 2019), and SCINA (Z. Zhang et al. 2019) were developed, which leverage broader cell type-specific marker gene data (X. Zhang et al. 2019) to probabilistically assign single cells to cell types specified in terms of marker genes or an 'unknown' category, in the absence of clustering. While these methods perform well when distinguishing between broad cell type lineages, their performance can be limited when distinguishing more closely related cell types, such as regulatory and memory T cells (Grabski & Irizarry 2020).

A further set of methods have been developed to leverage scRNA-seq to link tumour genotype to phenotype (in this case gene expression), both when the genotype is inferred directly from the gene expression or measured using an orthogonal assay. In the former case, the dosage dependence of gene expression on copy number (Venteicher et al. 2017; Müller et al. 2018) means scRNA-seq data can be leveraged to predict large-scale *copy number aberrations*, allowing for approximate inference of clonal composition and separation of malignant cells from putatively normal diploid cells. One class of methods to do this infers copy number profiles directly from scRNA-seq data by comparing expression profiles of input cells to a background constructed from normal cells (Fan et al. 2018; Serin Harmanci et al. 2020). These methods rely on large contiguous regions of up- or down-regulation relative to background that correspond to copy number-dependent changes in expression rather than alterations in gene regulation. CONICSmat (Müller et al. 2018) removes the reliance on a normal (non-tumour) set of cells specified *a priori* by fitting a bimodal Gaussian to expression values derived from each segment of the genome, but user interpretation of the results is required to distinguish putative non-malignant and malignant cells. CaSpER (Serin Harmanci et al. 2020) additionally incorporates allelic frequency data to correct CNV calls by identifying regions of LOH. All of these methods depend on the reliability of the specified or inferred background cells as a baseline for the euploid state — background cells derived from a different tissue source or patient may lead to over- or under-calling. Alternatively, when ground truth copy number profiles are available from orthogonal data types, such as single-cell whole-genome sequencing, scRNA-seq profiles can be directly aligned to these with CloneAlign (Campbell et al. 2019). Somatic SNV information derived from bulk or single-cell DNA-seq data, instead of CNVs, can also be used to align scRNA-seq data (McCarthy et al. 2020). In addition to inferring large-scale clonal structure, these methods can also be useful for distinguishing malignant cells from nonmalignant tissue-related cell types when established marker genes are unavailable.

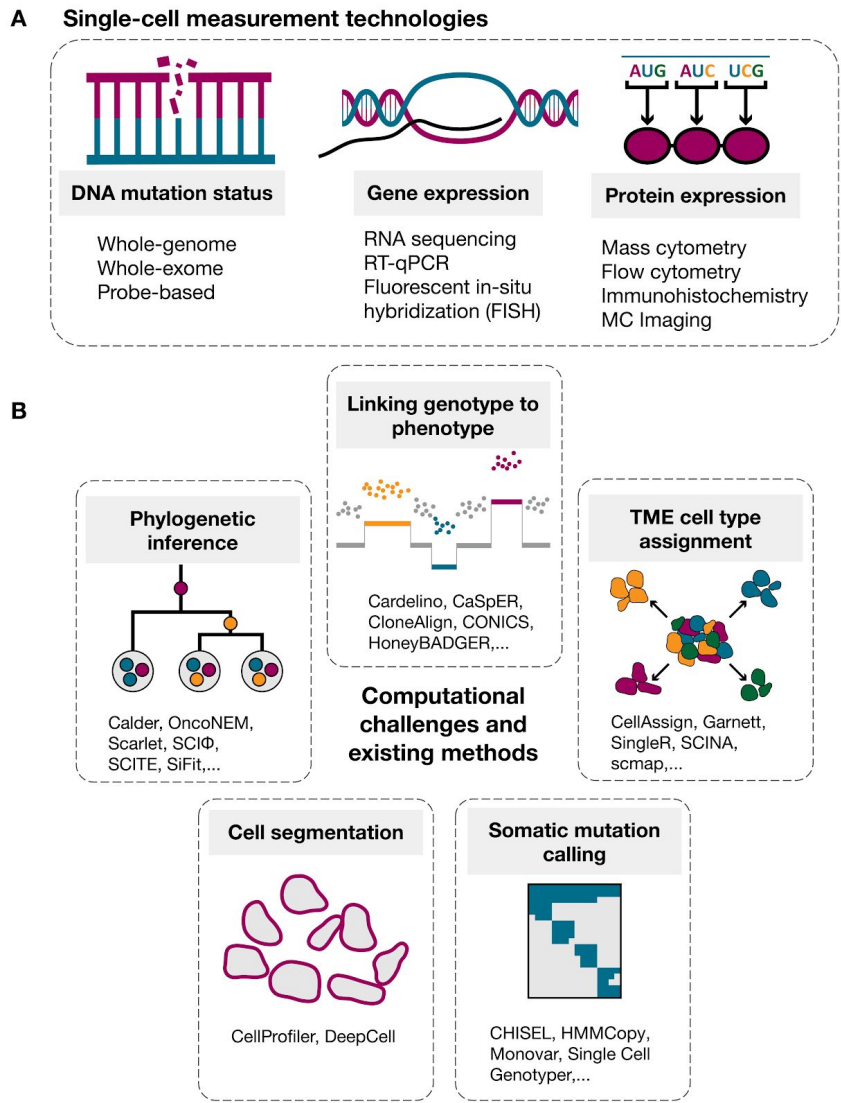

**Figure 1:** technologies and computational methods for single-cell cancer genomics. **A** Experimental technologies frequently used in cancer genomics for profiling single-cells discussed in this review. **B** Examples of common computational challenges and current solutions in the analysis of single-cell cancer genomic data.

A further promising technological advance in gene expression profiling of tumour cells concerns whole-transcriptome profiling of tissues and single-cell or near-single-cell resolution with the retention of information on the spatial origin of the region profiled. The main technologies in this area — such as the 10X Genomics Visium system — achieve a resolution of 100μm (3-30 cells), with the state-of-the-art attaining 2μm resolution enabling sub-cellular measurements

(Vickovic et al. 2019). To interpret these data, methods that attempt to understand the spatial distribution of cell types and states in the tumour microenvironment, either through *de novo* discovery or integration with disaggregated scRNA-seq, have been developed. These include Multimodal Intersection Analysis (MIA) (Moncada et al. 2020) which quantifies the overlap between expression cluster- and region-specific marker genes using a hypergeometric statistical test, and Spatial

Transcriptome Deconvolution (STD, Maaskola et al. 2018) which employs a hierarchical Bayesian model to locally deconvolve measured spots into "transcriptomic factors" that serve as a lower dimensional representation of the data.

## Protein expression

Mass cytometry approaches (see Spitzer & Nolan 2016 for an overview) extend classical immunohistochemistry to simultaneously interrogate up to 40 distinct markers using heavy metal-tagged antibodies directed to cell surface, cytoplasmic, and nuclear epitopes, at single-cell resolution. Like immunohistochemistry, these techniques can be applied to archival (fixed) material, which is readily available and typically not suitable for single-cell genome or transcriptome characterization. Compared to single-cell RNA-sequencing, mass cytometry costs substantially less per cell and allows users to focus on biologically meaningful proteins rather than primarily highly expressed genes (Spitzer & Nolan 2016), though with the obvious drawback of not capturing full-transcriptome data. There have been many applications of mass cytometry in cancer research, ranging from profiling apoptotic pathways to identifying putative drug targets (Teh et al. 2020), building predictive models of patient relapse at diagnosis (Good et al. 2018), and quantifying of immune cell infiltration across large patient cohorts (Wagner et al. 2019).

Few tools tailored for the analysis of mass cytometry cancer datasets currently exist, though many existing tools can be repurposed to answer cancer-specific questions (see Nowicka et al. 2017 for a recommended cytometry workflow). Normalization and signal correction can be applied using tools such as CATALYST (Chevrier et al. 2018), which employs bead-based approaches to account for signal spillover due to contaminating channels. A common secondary step is cell type identification, typically performed via clustering algorithms including bespoke models for mass cytometry such as Phenograph (Levine et al. 2015). Supervised marker gene-based approaches, comparable to CellAssign or Garnett, have yet to be tested in mass cytometry data. Finally, methods such as SPADE (Qiu et al. 2011) allow for the ordering of cell populations based on gradual changes in protein expression, with successful applications identifying cellular transitions in breast cancers (Giesen et al. 2014).

Recently, multiple mass cytometry-based methods that measure the cellular proteome at single cell-level or subcellular resolution that retain spatial information of cells *in situ* have been developed. These methods, collectively referred to as mass cytometry imaging (MCI) methods, include imaging mass cytometry (IMC, Giesen et al. 2014), which can profile to a spatial resolution of 1um, and multiplexed ion beam imaging (MIBI, Angelo et al. 2014), which can profile to 200-300nm. These methods provide new insights into homotypic and heterotypic cell-to-cell interactions, immune infiltration, and physical microenvironment architecture that cannot be inferred from disaggregated data alone.

The analysis of MCI data consists of 4 major steps: (1) cell segmentation, (2) normalization, (3) cell type assignment, and (4) spatial analysis. As steps (2) and (3) use no specific spatial information they largely follow the process for disaggregated mass cytometry data as described above. Multiple methods exist for segmentation, including CellProfiler (McQuin et al. 2018), which implements a two-step approach, using markers to first establish nuclear boundaries and then predict cellular boundaries, which are typically more variable in size and shape. Additional approaches such as Ilastik (Berg et al. 2019) and DeepCell (Van Valen et al. 2016) are supervised approaches that can be used to establish cellular boundaries based on pixel classification by leveraging random forests and convolutional neural networks, respectively.

Currently, there are few computational methods that integrate the available spatial and phenotypic information to provide insights into cell-cell interactions or tumour architecture, with most studies to date employing *ad hoc* methods to assess broad patterns in spatial proximity and composition. For example, (Jackson et al. 2020) use a Louvain algorithm to find clusters of neighbouring cells ('spatial communities') and characterize these communities according to cell type abundance. Similarly, Keren et al. 2018 used proportions of co-localizing tumour and immune cells to define three archetypal communities: cold, compartmentalized, and mixed. Initial analyses (Keren et al. 2018, Jackson et al. 2020) show that this type of spatial information may improve prediction of patient survival based on cell type abundance. When larger panels of proteins can be simultaneously profiled at the single-cell level, methods that leverage existing receptor-ligand databases such as CellPhoneDB (Efremova et al. 2020) may provide insights into context-dependent cell-cell interactions in the tumour microenvironment.

## Unsolved problems and future directions

There has been a remarkable drive by the research community to create computational methods for the analysis of single-cell data, with over 600 tools at time of writing created for scRNA-seq data alone (Zappia et al. 2018). However, tools specifically designed for the analysis of single-cell cancer genomic data are still in their infancy. Here, we outline future computational methods across four research domains (summarized in **Table 2**) that we envisage as necessary to answer pressing questions about cancer initiation, progression, and resistance from single-cell data.

### Uncovering clonal dynamics from single-cell genomic data

Interpreting mutational dynamics at the single-cell level remains one of the major challenges in single-cell genomics. While the robustness of single-cell copy number technologies such as  direct library preparation (DLP, Zahn et al. 2017) has been recently demonstrated and substantial progress has been made in calling CNVs from these data (Dong et al. 2019; Wang et al. 2019), little attention has been paid to the problem of constructing phylogenetic trees from single-cell CNV data despite the large number of such methods for point mutation data. Furthermore, systematic methods for cutting single-cell phylogenies   to identify *clones* are lacking, with most studies to date employing phylogeny-naive approaches such as hierarchical clustering (Zahn et al. 2017) or density-based clustering (Laks et al. 2019). Existing methods developed for clone-level phylogeny construction from bulk genomic data (Malikic et al. 2015) cannot currently be deployed on single-cell data due to computational time constraints. The lack of bespoke methods in this space is notable given the extremely large number of comparable methods for clustering scRNA-seq data, though we expect this gap to be filled as commercial platforms for single-cell CNV sequencing become widely available.

With an increasing number of multi-sample cancer datasets being generated, we expect that approaches to understand and reconstruct clonal dynamics from timeseries or spatially-sampled single-cell mutation data will be developed. Single-cell whole genome data is the ideal substrate for  evaluating clonal fitness in the treatment-naïve context and after chemotherapeutic intervention, either from longitudinally-collected samples or from cross-sectional data exploiting topological signatures of fitness in phylogenetic trees. Furthermore, the problem of reconstructing clonal migration histories is particularly important, with early studies demonstrating the timing of genomic mutations relative to clonal invasion (Casasent et al. 2018). Such methods operating at the single-cell level could leverage phylogenetic placement and would complement those for establishing clonal migration patterns from bulk DNA sequencing data such as MACHINA (El-Kebir et al. 2018).

### Impact of the TME on tumour evolution and phenotypes

As computational tools mature to provide quantitative estimates of clonal fitness from single-cell genome sequencing technologies and similar tools emerge to quantify tumour phenotypic states and microenvironment composition from single-cell expression profiling, there is a major need for the development of methodologies to interrogate the interplay between these two important facets. For example, it is incompletely understood how the composition of the microenvironment shapes the clonal fitness landscape and preferentially allows for the growth of certain tumours. We envisage in the future that given sufficient data there will exist computational tools to make quantitative predictions of clonal fitness given clonal genotypes and possibly phenotypes, in response to perturbations to the local microenvironment composition and expression phenotype. Such models are crucial to enable an era of personalized chemotherapy, where therapeutic interventions will be actioned not only on the basis of (sub-) clonal genotypes, but also on possible interactions with the TME. These tools will be further enabled by advances in immune-related sequencing technologies, such as the ability to sequence T- and B-cell receptor sequences and genotype the HLA loci responsible for MHC class I antigen presentation. The combination of *in situ* sequencing chemistry with these technologies would allow for direct interrogation of the T- and B-cell clonotypes that are spatially adjacent and likely respond to individual tumour clones.

## Machine learning and biomarker-based methods to guide therapy

Arguably one of the major goals of modern cancer research is "personalized chemotherapy" — the ability to tailor therapies based on patient-specific characteristics with the goal of achieving better outcomes than treatment with histotype-based standard-of-care. Single-cell technologies can enable this goal in two ways: (i) the identification of rare cell populations that are likely to form resistant clones and lead to patient relapse, and (ii) the association of cell type-specific phenotypic states (either in the tumour itself or the TME) that are associated with particular clinical courses and therapy responses.

Realizing these scenarios requires research discovery and clinical implementation phases with computational tools needed to support both. For rare cell population identification, multiple tools exist for scRNA-seq data (Wegmann et al. 2019; Jindal et al. 2018) but further interpretation of whether such rare populations have proliferative or chemoresistant potential will require the development of new methods that integrate multiple distinct modalities such as timeseries transcriptomic and genomic data. Translating these inferences to the clinical context will demand methods that can quickly and robustly identify such populations while accounting for multiple levels of technical variability including substantial inter-patient heterogeneity. For the discovery phase of the association of cell-type specific phenotypic states to outcomes, a current limiting factor is the availability of cohort-scale single-cell datasets with follow-up treatment and outcome data, on top of which existing tools such as CellAssign (A. W. Zhang et al. 2019) and edgeR

(Robinson et al. 2010) could be deployed. In the clinical implementation phase, the design of robust probabilistic machine learning models is necessary to predict outcomes with calibrated uncertainty, taking into consideration the ability to easily *overfit* to the training datasets given their high dimensional nature and relatively low number of samples.

## Exploiting spatial data to model tissue dynamics

The advent of technologies to profile tumour cells while retaining information about their spatial origin has provided the opportunity to uncover mechanisms crucial to tumour initiation and progression, such as cell-to-cell signalling and tumour infiltration. However, the handful of papers in this field approach these computational analyses in an ad-hoc manner, and thus we expect there to be a number of efforts to systematically address these questions. Firstly, methods are needed to quantitatively infer cell-to-cell signalling networks from spatially resolved data; this may take the form of simulation-based inference using techniques such as *Approximate Bayesian Computation* due to the fundamental difficulty of assessing whether a cell exists in a phenotypic state due to signalling from a proximal cell, or vice versa. Secondly, we expect the development of methods to integrate data from disaggregated whole-transcriptome technologies such as scRNA-seq with panel-based technologies such as IMC. Thirdly, there is a need for methods to automate the descriptions of spatially resolved datasets which can have cohorts of hundreds or thousands of patients, identifying regions of immune cell infiltration in tumours or regions of high proliferation in a systematic manner across patients. Fourthly, as the majority of genomic assays that retain spatial information operate in 2D only, we anticipate the development of methods to reconstruct 3D expression profiles through the integration of serial 2D "slices". Finally, as the data produced by such technologies are shown to be predictive of patient outcomes, the need for machine learning models that the irregularly-sampled high-dimensional spatial data as input will become necessary.

| Research domain/modality | Computational tools necessary |
|---|---|
| Low-depth single-cell whole genome sequencing | • Copy number aware clustering of scDNA-seq (clonal inference)<br>• Phylogeny tree reconstruction from copy number calls |
| Tumour-microenvironment interactions | • Simulation-based or statistical modelling tools to asses impact of TME phenotypes and composition on clonal evolution |
| Single-cell biomarker approaches | • Quantification of proliferative and resistance potential of rare cell populations<br>• Robust probabilistic machine learning models to predict outcomes from gene expression phenotypes and TME composition |
| Spatial tumour dynamics | • Inference of cell-cell signalling networks<br>• Integration of spatially-resolved data with whole transcriptome scRNA-seq<br>• Automated cell type inference<br>• Reconstruction of 3D tumour architecture from 2D measurements |

**Table 2:** future research domains and possible solutions for single-cell computational genomics.

## Discussion

In this review, we have outlined the existing computational models designed for the analysis of single-cell genomic, transcriptomic, and proteomic data in the context of cancer research, and described four emerging domains in which the development of future computational tools is essential. However, there exist additional measurement modalities and analysis strategies that will be equally crucial to unlocking insights from single-cell cancer data.

In particular, the generation and analysis of single-cell epigenetic data — including scATAC-seq for accessibility (see Chen et al. 2019 for a computational methods overview), scCHIP-seq for various types of histone modifications (Grosselin et al. 2019), single-cell bisulfite sequencing (Smallwood et al. 2014) for 5-methylcytosine marks, and scAba-seq (Mooijman et al. 2016) for 5-hydroxymethylcytosine modifications — represents a promising novel modality for tackling fundamental questions related to cancer heterogeneity and resistance. For example, in an application of high-throughput scCHIP-seq to breast cancer to breast cancer patient-derived xenografts, (Grosselin et al. 2019) found a low-prevalence population of cells in a pre-treatment sensitive tumour that shared chromatin features with all post-treatment resistant cells, which was not observable at a bulk level.

However, the analysis of such data remains in its infancy, with few bespoke computational methods for the analysis of cancer single-cell epigenomic data, to our knowledge. Furthermore, it is not apparent that the development of such tools will be necessary if standard workflows for resolving cellular heterogeneity and identification of differential regions of epigenetic marks are sufficient to answer questions in a cancer-specific context.

A similarly important emerging area of research concerns the development of technologies that measures multiple data modalities simultaneously — such as G&T-seq (Macaulay et al. 2015) for the combined measurement of the genome and transcriptome at single-cell resolution, and scNMT-seq (Clark et al. 2018) for DNA methylation, chromatin accessibility, and transcription. For example, the ability to jointly measure the genome and transcriptome at single-cell level across many tumour cells would enable the refined tracking of expression changes linked to ongoing clonal evolution, tying possibly targetable phenotypes to waves of clonal expansion. There exist bespoke computational tools for the joint analysis of bulk multi-omic data generated through efforts such as The Cancer Genome Atlas (Weinstein et al. 2013), including latent factor models such as Multi Omic Factor Analysis (Argelaguet et al. 2018) and DIABLO (Singh et al. 2019). However, to date few methods for integrating multi-omic data at the single-cell level in a cancer context exist, and the extent to which such methods will be necessary remains to be seen. For example, in the former example of G&T-seq applied to tumour cells, one could define clones using one of the previously mentioned computational methods, then identify transcriptional signatures of such clones using the many differential expression methods designed for (single-cell) RNA-sequencing.

Finally, a crucial area of computational methods development that permeates cancer research concerns the visualization of single-cell data. For example, E-scape (Smith et al. 2017) contains a suite of methods for the visualization of single-cell mutation data, including *CellScape* for visualizing single-cell copy number heatmaps and phylogenies, and *TimeScape* for visualizing clonal prevalence over time. Meanwhile, some existing tools have visualization capabilities built-in such as CaSpER (Serin Harmanci et al. 2020), while other methods such as Millefy (Ozaki et al. 2020) will be exceptionally useful for assessing variant detection from scRNA-seq despite not being designed specifically in a cancer context.

## Acknowledgements

We thank Sally Millett for the creation of the visualizations in Figure 1. We acknowledge the support of the Natural Sciences and Engineering Research Council of Canada (NSERC).